\documentclass[pra,preprint,showpacs]{revtex4}
\usepackage[latin1]{inputenc}
\usepackage{eufrak}
\usepackage{dcolumn}
\usepackage{amsmath}
\usepackage{color}

\usepackage[dvips]{graphicx}

\newcommand{\phial}{\phi_{\alpha}}

\newcommand{\phibe}{\phi_{\beta}}

\newcommand{\albe}{{\alpha\beta}}

\newcommand{\al}{\alpha}
\newcommand{\be}{\beta}
\newcommand{\ga}{\gamma}

\newcommand{\pdt}{\frac{\partial}{\partial t}}
\newcommand{\ddt}{\frac{\text{d}}{\text{d}t}}
\newcommand{\rpsi}{| \Psi \rangle}

\newcommand{\bega}{{\beta\gamma}}

\newcommand{\R}{\vec{R}}
\newcommand{\rr}{\vec{r}}

\newcommand{\Hbega}{H_{\beta\gamma}}
\newcommand{\Bbega}{B_{\beta\gamma}}
\newcommand{\bra}[3]{\left\langle #1 \left | #2 \right | #3 \right\rangle}
\newcommand{\braphi}[1]{\bra{\phial}{#1}{\phibe}}
\newcommand{\braa}[2]{\left\langle #1 \left | #2 \right . \right\rangle}

\newcommand{\Halbe}{H_{\albe}}
\newcommand{\Balbe}{B_{\albe}}
\newcommand{\Talbe}{T_{\albe}}

\newcommand{\rmi}{\text{i}}

\newcounter{fusszaehler}

\newcommand{\mittig}[1]{\begin{center}#1\end{center}}

\begin{document}

\title{Non-adiabatic quantum molecular dynamics: Ionization of many electron systems}
\author{Mathias Uhlmann}
\email{Mathias.Uhlmann@gmx.de}
\homepage{www.dymol.org}
\author{Thomas Kunert}
\author{R\"udiger Schmidt}
\email{schmidt@physik.tu-dresden.de}
\affiliation{Institut f\"ur Theoretische Physik\\Technische Universit\"at Dresden\\D-01062 Dresden\\Germany}

\date{\today}

\begin{abstract}
We propose a novel method to describe realistically ionization processes with absorbing boundary conditions in basis expansion
within the formalism of the so-called Non-Adiabatic Quantum Molecular Dynamics.
This theory  couples self-consistently a classical description of the nuclei with a quantum mechanical treatment of the electrons in
atomic many-body systems.
In this paper we extend the formalism by introducing absorbing boundary conditions
 via an imaginary potential. It is shown how this potential can be constructed in
 time-dependent density functional theory in basis expansion.
The approach is first tested on the hydrogen atom and the pre-aligned hydrogen molecular ion H$_2^+$ in intense laser fields
where reference calculations are available. It is then applied to study the
ionization of non-aligned H$_2^+$ and H$_2$. 
Striking differences in the orientation dependence between both molecules are found.
Surprisingly, enhanced ionization is predicted for perpendicularly aligned molecules.
\end{abstract}


\maketitle

\section{Introduction}
\label{sec:intro}
The experimental and theoretical investigation of the interaction of atoms, molecules and clusters with intense laser fields represents one of the
most challenging problems of current research. 
In atoms, high harmonic generation (HHG)\cite{McPherson_JOSAB-4_595(1987),Ferray_JPB-21_L31(1988),Li_PRA-39_5751(1989)},
above threshold ionization \cite{Agostini_PRL-42_1127(1979),Kruit_PRA-28_248(1983),Fedorov_Atombuch} or
stabilization against ionization \cite{Pont_PRL_65_2362(1990),deBoer_PRL_71_3263(1993),deBoer_PRA_50_4085(1994),Fedorov_Atombuch} have been observed.
In molecules, due to the additional nuclear degrees of freedom (DOF), further mechanisms occur,
like molecular stabilization against dissociation \cite{Yuan-JCP_68_3040(1978),Bandrauk-JCP_74_1110(1981),Posthumus-JPB_33_L563(2000)},
bond softening \cite{Yao-PRA_48_485(1993),Charron-PRL_71_692(1993),
Giusti-Suzor-JPB_28_309(1995),Saendig-PRL_85_4876(2000),Williams-JPB_33_2743(2000)},
above threshold dissociation
\cite{Giusti-Suzor-PRL_64_515(1990),Bucksbaum-PRL_64_1883(1990),Williams-JPB_33_2743(2000)},
charge resonance enhanced ionization \cite{Codling-JPB_26_783(1993),Zuo-PRA_52_R2511(1995),Seideman-PRL_75_2819(1995),Williams-JPB_33_2743(2000)},
orientation dependent HHG~\cite{Yu-JCP_102_1257(1995),Kopold-PRA_58_4022(1998),Lappas-JPB_33_4679(2000),
Lein-PRL_88_183903(2002),Lein-PRA_66_023805(2002),Lein-PRA_67_023819(2003),Lein-PRA_66_023805(2002),Kamta-PRA_70_011404(2004),
Zimmermann-PRA_71_033401(2005),Chirila-submitted_2005},
molecular species dependent orientation dependence of single ionization in dimers~\cite{Litvinyuk-PRL_90_233003(2003),Alnaser-PRL_93_113003(2004),Zhao-PRA_67_043404(2003)}
or an
unexpected suppression of ionization in dimers in comparison to the so-called companion atom~\cite{Talebpour-JPB_31_2769(1998),Wells-PRA_66_013409(2002),
Talebpour-JPB_29_L677(1996),Guo-PRA_58_R4271(1998),DeWitt-PRL_87_153001(2001),Faisal-LP_9_115(1999),Muth-Boehm-PRL_85_2280(2000),Kjeldsen-JPB_37_2033(2004),
Dundas-PRA_71_013421(2005)}, to name but a few effects.

The theoretical understanding of these mechanisms requires, in principle, the solution of the
time-dependent Schr\"odinger equation (TDSE) for all electrons and all nuclear DOF. 
However, only for the smallest systems, 
atomic hydrogen~\cite{Geltman_JPB_33_1967(2000),Hansen_PRA_64_033418(2001)},
atomic helium~\cite{Dundas_JPB_32_L231(1999),Parker_JPB_33_L691(2000)},
laser aligned H$_2^+$~\cite{Chelkowski-PRA_52_2977(1995)} and
laser aligned H$_2$ with fixed nuclei~\cite{Harumiya-JCP_113_8953(2000),Harumiya-PRA_66_043403(2002)}
numerical solutions of the TDSE exist. 
In larger systems (or for hydrogen molecules without constraints) approximations are necessary 
due to the exponential scaling of the computational effort
with the number of DOF.

\nocite{Runge-PRL_52_997(1984)}
\nocite{Saalmann-ZPD_38_153(1996)}
\nocite{KS03}
\nocite{Uhlmann-Diplompaper}
\nocite{Calvayrac_JPB-31_5023(1998),Suraud_PRL-85_2296(2000),Calvayrac_PR-337_493(2000)}
\nocite{Dundas-JPB_37_2883(2004)}
\nocite{Castro-EPJD_28_211(2004)}
One possibilty consists in the combination of time-dependent density functional theory~\cite{Runge-PRL_52_997(1984)} (TDDFT)
and a classical description of the nuclei~\cite{Saalmann-ZPD_38_153(1996),KS03,Uhlmann-Diplompaper,Calvayrac_JPB-31_5023(1998),Suraud_PRL-85_2296(2000),Calvayrac_PR-337_493(2000),Dundas-JPB_37_2883(2004),Castro-EPJD_28_211(2004)}. 
Most of the approaches use a representation of the Kohn-Sham (KS) functions
on a grid to solve the time-dependent KS equations~\cite{Calvayrac_JPB-31_5023(1998),Suraud_PRL-85_2296(2000),Calvayrac_PR-337_493(2000),
Castro-EPJD_28_211(2004),Dundas-JPB_37_2883(2004)}. The first of these grid based approaches has been developed by Reinhard et al.~\cite{Calvayrac_JPB-31_5023(1998),Suraud_PRL-85_2296(2000),Calvayrac_PR-337_493(2000)}. More recently, Dundas~\cite{Dundas-JPB_37_2883(2004)}
and Castro et al.~\cite{Castro-EPJD_28_211(2004)} have also developed such methods. 

In contrast, a basis expansion of the KS-orbitals with local basis functions is used in the so-called non-adiabatic quantum molecular dynamics (NA-QMD),
developed in our group~\cite{Saalmann-ZPD_38_153(1996),KS03}.
It has been successfully applied so far to very different non-adiabatic processes, like atom-cluster collisions~\cite{Saalmann-PRL_80_3213(1998)},
ion-fullerene collisions~\cite{Kunert-PRL_86_5258(2001)}, 
laser induced excitation and fragmentation of molecules~\cite{Uhlmann-Diplompaper} or fragmentation and isomerization of organic molecules in laser 
fields~\cite{Kunert-Ethylen}. However, a realistic description of ionization with the NA-QMD theory is still an open problem.

In this work, we present a general method in basis expansion and extend the NA-QMD formalism to describe ionization in many-electron systems. To this end, two problems
have to be addressed. First, an appropriate basis set suitable for the description of highly excited and ionized states has to be found. We have focused
on this problem in a recent paper~\cite{BE1}. The second problem concerns the introduction of absorbing boundary conditions. This is still a challenging problem
for many-electron calculations performed on grids (see e.g.~\cite{Dundas-PRA_71_013421(2005)}), and in particular, completely open for methods using basis expansion.

In~\cite{Uhlmann-Diplompaper} we have used an ad-hoc manipulation
of the electronic expansion coefficients in one electron calculations. In this paper, we introduce general absorbing boundary conditions in basis expansion
via an imaginary potential. It will be shown, how such an potential can be build using the many-body Schr\"odinger equation as well as 
TDDFT.

The outline of this paper is as follows. First, we introduce absorbing boundary conditions in basis expansion for the general case of
the Schr\"odinger equation in section~\ref{sec:abs_oe}. 
In section~\ref{sec:abs_me}, the extension of the NA-QMD formalism including absorbing boundary conditions is presented. Details of the used absorbing potential
are outlined in section~\ref{sec:para}. The method is tested on the hydrogen atom (section~\ref{sec:H}) as well as aligned H$_2^+$ (section~\ref{sec:H2})
where reference calculations are available. In section~\ref{sec:H2_2}, the calculated ionization probabilities and rates of non-aligned H$_2^+$ and H$_2$ are presented.
A completely different orientation dependence between both molecules is found. In addition, the calculations predict surprisingly enhanced ionization
for \emph{perpendicularly} aligned molecules.

\section{Absorbing boundary conditions}

\subsection{The Schr\"odinger equation and basic idea}
\label{sec:abs_oe}

We start with the introduction of absorbing boundary conditions for the general case of the Schr\"odinger equation
 (atomic units are used throughout the paper)
\begin{equation}
	\rmi \pdt \rpsi  = \hat{H}(t) \rpsi
	\label{eq:SGL}
\end{equation}
where the Hamiltonian 
\begin{equation}
	\hat{H}(t) = \hat{T} + \hat{V}(t)
\end{equation}
consists of the operator for the kinetic energy $\hat{T}$ and the potential $\hat{V}$ which includes the two-body interaction. 
Thus, $\rpsi$ represents the, in general, many-particle wave function.
The absorbing boundary conditions are incorporated via an imaginary potential. The Hamiltonian is thus modified,
\begin{equation}
	\hat{H}_\text{abs}(t) = \hat{H}(t) - \text{i} \hat{V}_\text{abs}(t)
	\label{eq:hamabs}
\end{equation}
where $\hat{V}_\text{abs}$ is the absorbing potential.
With a hermitian $\hat{V}_\text{abs}$ such an operator $\hat{H}_\text{abs}$ is not hermitian and therefore the norm is not conserved, i.e.
\begin{eqnarray}
	\ddt \langle \Psi | \Psi \rangle &=&
	\langle \Psi | \rmi\hat{H}_\text{abs}^* - \rmi\hat{H}_\text{abs} | \Psi \rangle \nonumber\\
	&=& -2 \langle \Psi | \hat{V}_\text{abs} | \Psi \rangle \,\text{.}
	\label{eq:normaenderung}
\end{eqnarray}
A semi-positive definite $\hat{V}_\text{abs}$ leads to the \emph{desired} effect of norm reduction, i.e. the derivative in~(\ref{eq:normaenderung}) is 
negative or zero. 

The balance of the total energy 
\begin{equation}
	E(t) = \langle \Psi(t) | \hat{H}(t) | \Psi(t) \rangle
\end{equation}
is changed if the absorber potential is used in the propagation of $|\Psi\rangle$, i.e.
\begin{equation}
	\ddt E = \langle \Psi | \ddt\hat{V} | \Psi \rangle + \Delta_\text{abs}
	\label{eq:eb_allg}
\end{equation}
with 
\begin{equation}
	\Delta_\text{abs} = - \langle \Psi | \hat{V}_\text{abs}\hat{H} + \hat{H}\hat{V}_\text{abs} | \Psi \rangle \,.
	\label{eq:deltaabs_allg}
\end{equation}
The additional term is due to the absorber potential and changes the energy balance. 
Its actual effect depends on the definition of the potential $\hat{V}_\text{abs}$.

Introducing the time-dependent eigenfunctions $|\chi_a\rangle$ to $\hat{H}$
\begin{equation}
	\hat{H}(t) |\chi_a(t)\rangle = E_a( t ) |\chi_a(t)\rangle
	\label{eq:tdev_sgl}
\end{equation}
we define the absorbing potential as
\begin{equation}
	\hat{V}_\text{abs} = \sum_{a=1}^{\infty} f_a |\chi_a\rangle \langle \chi_a | \,\text{.}
	\label{eq:vabs_allg}
\end{equation}
The states $|\chi_a\rangle$, sometimes called ``field-following'' adiabatic states~\cite{Kono-LP_13_883(2003),Kono-CP_304_203(2004)}, form an
orthonormal set.
With our definition of $\hat{V}_\text{abs}$ these states $|\chi_a\rangle$ are also eigenstates to $\hat{H}_\text{abs}$
\begin{equation}
	\hat{H}_\text{abs}(t) |\chi_a(t)\rangle = ( E_a( t ) - \rmi f_a ) |\chi_a(t)\rangle
\end{equation}
but lead to imaginary eigenenergies, i.e. finite lifetimes.
The factors $f_a$ determine the strength 
of the absorber at a certain energy and are discussed in section~\ref{sec:para}.
The wave function $|\Psi(t)\rangle$ is now expanded also in these eigenfunctions
\begin{equation}
	| \Psi \rangle = \sum_{a=1}^{\infty} a_a | \chi_a \rangle \,.
	\label{eq:entw_dia}
\end{equation}
Inserted into the time-derivative of the norm~(\ref{eq:normaenderung}) this yields
\begin{equation}
	\ddt \langle \Psi | \Psi \rangle =
	-2 \sum_{a=1}^{\infty} \left|a_a\right|^2 f_a 
	\label{eq:normaenderung2}
\end{equation} 
and the additional term $\Delta_\text{abs}$~(\ref{eq:deltaabs_allg}) of the energy balance 
becomes
\begin{equation}
	\Delta_\text{abs} = -2  \sum_{a=1}^{\infty} \left|a_a\right|^2 f_a E_a \,\text{.}
	\label{eq:deltaabs_allg2}
\end{equation}
From equation~(\ref{eq:normaenderung2}), one can see that
the absorber potential decreases the norm of arbitrary wave functions $|\Psi \rangle$ only, if all $f_a\geq0$. 
Furthermore, it has to be guaranteed that electronic density in bound states is not affected by the absorbing potential.
In calculations on spatial grids this is approximately satisfied by applying the absorbing boundary
conditions 
far away from the nuclei (see e.g.~\cite{Neuhasuer-JCP_90_4351(1989),Chelkowski-PRA_52_2977(1995),Dundas-JPB_33_3261(2000)}). 
In our case of basis expansion~(\ref{eq:entw_dia}), this condition can naturally be fulfilled if the time-dependent eigenvalues $f_a$ of
$\hat{V}_\text{abs}$~(\ref{eq:vabs_allg}) are chosen to be
\begin{equation}
	f_a = \left\{ 
		\begin{array}{ccc}
			f_a = 0 & \hspace{0.5cm} \text{if} \hspace{0.5cm} & E_a \leq0 \\
			f_a > 0 & \text{if} & E_a > 0
		\end{array}
	\right. \,,
	\label{eq:f_a}
\end{equation}
i.e. the absorbing potential acts only on states in the continuum.
Thus, $\Delta_\text{abs}$ is always zero or negative if the potential is defined as in~(\ref{eq:vabs_allg}) and if the $f_a$ meet the criterion~(\ref{eq:f_a}).

\subsection{Non-Adiabatic Quantum Molecular Dynamics}
\label{sec:abs_me}

So far we have shown how to introduce absorbing boundary conditions if the Schr\"odinger equation in basis expansion is used. 
We show in the following how to introduce an absorbing potential within the NA-QMD formalism.

The coupled equations of motion (EOM) for the nuclear and electronic system have been given elsewhere in TDDFT~\cite{KS03} and time-dependent 
Hartree-Fock (TDHF)~\cite{Kunert-Ethylen} and will not be repeated here. Instead
we will present the changes in the single-particle EOM arising from an additional imaginary potential in the effective single-particle potential.

The single-particle wave functions are expanded in a local basis
\begin{equation} 
	\label{Basis} \Psi^{j\sigma}(\rr,t)=\sum_{\alpha=1}^{N_\text{b}}
	a^{j\sigma}_{\alpha}(t)\phi_\alpha(\rr-\R_{A_\alpha}(t)) \hspace{1cm} \text{with} \hspace{0.5cm} j=1\dots N_\text{e}^\sigma,\,\sigma=\uparrow,\downarrow
\end{equation} 
and only the expansion
coefficients $a^{j\sigma}_{\alpha}(t)$ are explicitly time-dependent. The symbol $A_\alpha$ denotes the atom
to which the atomic basis function $\phial$ is attached.
The $N_\text{b}$ basis functions are
either located at the nuclei, which in general move, or are located at fixed positions in space~\cite{BE1}. 
$N_\text{e}^\sigma$ is the number of electrons with the spin $\sigma$.

The variation of the total action with respect to electronic expansion coefficients and nuclear coordinates leads to self-consistently 
coupled EOM~\cite{KS03,Kunert-Ethylen}
for the electronic expansion coefficients $a^{j\sigma}_{\alpha}(t)$ and for the nuclear coordinates $\R_A$.
Here we add an imaginary potential to the effective single-particle Hamiltonian. The EOM for the electronic expansion coefficients are then modified
(cf. with \cite{KS03,Kunert-Ethylen})
\begin{equation}
        \label{KS_abs_me}
        \ddt a^{j\sigma}_\alpha=-\sum_\bega^{N_\text{b}} \left(S^{-1}\right)_\albe \left(\rmi\Hbega^\sigma + V_{\text{abs,}\be\ga}^{\sigma} + \Bbega\right)
	a^{j\sigma}_\gamma \qquad j=1,\dots ,N_\text{e}^\sigma,\,\sigma=\uparrow,\downarrow \,\text{.}
\end{equation}
In (\ref{KS_abs_me})
\begin{equation}
  \label{Salbe}
   S_\albe = \braa{\phial}{\phibe}
\end{equation}
is the overlap matrix between basis states,
\begin{equation}
  \label{Halbe_eff}
  \Halbe^\sigma = \braphi{\hat{H}_\text{eff}^\sigma}
\end{equation}
is the effective Hamilton matrix
with $\hat{H}_\text{eff}^\sigma$ the effective one-particle Hamiltonian~\cite{KS03,Kunert-Ethylen},
\begin{equation}
        \Balbe=\braa{\phial}{\ddt\phibe}
	\label{Balbe}
\end{equation}
is the non-adiabatic coupling matrix
and 
\begin{equation}
	V_{\text{abs,}\albe}^{\sigma} = \braphi{\hat{V}_\text{abs}^\sigma}
	\label{Vabs_naqmd}
\end{equation}
is the matrix element of the additional absorber potential introduced here and still to be defined (see below).

At this point we note explicitely that the EOM~(\ref{KS_abs_me}) are exactly the same for TDDFT~\cite{KS03} and TDHF~\cite{Kunert-Ethylen}. The
difference between both approaches consists in the calculation of the matrix elements $H_\albe^\sigma$ which in case of TDHF contains the non-local exchange term.
Thus, the whole following discussions and derivations belong simultaneously to both approaches, TDDFT~\cite{KS03} and TDHF~\cite{Kunert-Ethylen}.

With the additional damping term~(\ref{Vabs_naqmd})
the time-dependence of the norm (i.e. the total number of electrons in this case) becomes
\begin{equation}
	\ddt N = \ddt \sum_{\sigma=\uparrow,\,\downarrow} \sum_{j=1}^{N_\text{e}^\sigma} \sum_{\albe=1}^{N_\text{b}} a_\al^{j\sigma*} a_\be^{j\sigma} S_\albe
	=  - 2\, \sum_{j\sigma\albe} V_{\text{abs,}\albe}^\sigma a_\al^{j\sigma*} a_\be^{j\sigma} \,.
	\label{eq:normaenderung_naqmd}
\end{equation}
The total energy reads~\cite{KS03}
\begin{eqnarray}
	E(t) &=& 
	U(\R,t) + \sum_{A=1}^{N_\text{i}} \frac{M_A}{2} \dot{\R}_A^2 
	+ \sum_{\sigma=\uparrow,\,\downarrow} \sum_{j=1}^{N_\text{e}^\sigma} \sum_\albe^{N_\text{b}} a_\al^{j\sigma*} T_\albe a_\be^{j\sigma} \nonumber\\
	&& + \int \text{d}^3r \rho(\rr,t) \left(
		V(\rr,\R,t) + \frac12 \int \frac{\rho(\rr\,',t)}{\left| \rr - \rr\,' \right|} \text{d}^3r'
	\right) + E_\text{xc}[\rho](t)
\end{eqnarray}
with the kinetic and potential energy $U(\R,t)$ of the nuclei, the external potential $V(\rr,\R,t)$ that contains the electron-nuclear 
interaction and e.g. a laser field $V_\text{L}(t)$,
the electronic density
\begin{equation}
  \label{Rho} \rho(\rr,t)=\sum_{\sigma=\uparrow,\downarrow} \sum_{j=1}^{N^\sigma}\Psi^{j\sigma*}(\rr,t) \Psi^{j\sigma}(\rr,t) \,,
\end{equation} 
the matrix element of the kinetic energy of the electrons
\begin{equation}
        \Talbe=\braphi{-\frac{\Delta}{2}}
\end{equation}
and the exchange and correlation energy $E_\text{xc}$.
With the EOM~(\ref{KS_abs_me}) and the classical EOM for the nuclei~\cite{KS03}, which are not changed by the imaginary potential, one obtains for the energy
balance
\begin{equation}
	\ddt E = \int \rho( \rr, t ) \frac{\partial V_\text{L}( \rr, t )}{\partial t} \text{d}^3r
	- \sum_{A=1}^{N_\text{i}} Z_A  \frac{\partial V_\text{L}( \R_A, t )}{\partial t}  + \Delta_\text{abs}
	\label{EB_abs_me}
\end{equation}
with $V_\text{L}$ the external, time-dependent potential (e.g. a laser) and the additional term
\begin{equation}
	\Delta_\text{abs} 
	= - \sum_{\sigma=\uparrow,\downarrow} \sum_{j=1}^{N_\text{e}^\sigma} \sum_{\albe\ga\delta}^{N_\text{b}}
	\left(
		V_{\text{abs, }\al\ga}^{\sigma} \left(S^{-1}\right)_{\ga\delta} H_{\delta\be}^\sigma
		+ H_{\al\ga}^\sigma \left(S^{-1}\right)_{\ga\delta} V_{\text{abs, }\delta\be}^{\sigma}
	\right) a_\al^{j\sigma*} a_\be^{j\sigma} 
	\label{Deltaabs}
	\,.
\end{equation}
The first two terms arise naturally from the interaction of the electronic density $\rho(\rr,t)$ and the nuclei (with charge $Z_A$) with an external field.
They are of course identical with the energy balance given in \cite{KS03}. The last term $\Delta_\text{abs}$ in~(\ref{EB_abs_me}) is evidently induced by
the imaginary potential. We note explicitely that the non-adiabatic coupling matrix $B_\albe$~(\ref{Balbe}), which at first glance
seems to act equivalently to the absorbing part $V_{\text{abs, }\albe}^\sigma$ in (\ref{KS_abs_me}),
does not affect the energy balance because these terms are
canceled out in the calculation of $\frac{\text{d}E}{\text{d}t}$ due to the classical EOM as it should be and as it has been shown in~\cite{KS03}.

The general results (\ref{eq:normaenderung_naqmd}) and (\ref{Deltaabs}) are equivalent to (\ref{eq:normaenderung}) and (\ref{eq:deltaabs_allg}).
They are valid for any absorbing potential $\hat{V}_\text{abs}^\sigma$ and any basis set $\{|\phi_\al\rangle\}$. The same holds for the EOM~(\ref{KS_abs_me}).
However, physically, any choice of $\hat{V}_\text{abs}^\sigma$ must guarantee that the absorption is only applied to that part of 
the density which belongs to the continuum.
This is equivalent to the requirement that norm and energy are decreased for any arbitrary density (cf. also with section~\ref{sec:abs_oe}), i.e. that
\begin{equation}
	\Delta_\text{abs}
	\,\leq\, 0 \hspace{0.5cm} \text{and} \hspace{0.5cm} \ddt N \leq 0 \hspace{1cm} \forall \, a_\al^{j\sigma} \, \text{.}
\end{equation}
In order to realize that we make use of the general idea presented in section~\ref{sec:abs_oe}.

To this end, the single-particle wave functions $|\Psi^{j\sigma}\rangle$ are expanded in the, now, effective single-particle ``field-following'' adiabatic states,
i.e. (cf. equation~(\ref{eq:entw_dia}))
\begin{equation}
	\label{Basis_diag} |\Psi^{j\sigma}\rangle(t)
	=\sum_{a=1}^{N_\text{b}} a^{j\sigma}_{a}(t)|\chi_a\rangle(t)) 
	\hspace{1cm} \text{with} \hspace{0.5cm} j=1\dots N_\text{e}^\sigma,\,\sigma=\uparrow,\downarrow \,.
\end{equation}
with $|\chi_a^\sigma(t)\rangle$ defined as (cf. equation~(\ref{eq:tdev_sgl}))
\begin{equation}
	\hat{H}_\text{eff}^\sigma(t) |\chi_a^\sigma\rangle(t) = \epsilon_a^\sigma(t)  |\chi_a^\sigma\rangle(t) \,.
	\label{eq:tdev_dft}
\end{equation}
The absorber potential is formally constructed as before (cf. equation~(\ref{eq:vabs_allg}))
\begin{equation}
	\hat{V}_\text{abs}^\sigma = \sum_{a=1}^{N_\text{b}} f_a^\sigma |\chi_a^\sigma\rangle \langle \chi_a^\sigma | \,.
	\label{eq:vabs2}
\end{equation}
In principle, the expansion coefficients $a_a^{j\sigma}$ in equation~(\ref{Basis_diag}) can be obtained 
from the EOM~(\ref{KS_abs_me}) written for the basis~(\ref{Basis_diag}).
In this case, the basis functions themselves would depend on the effective Hamiltonian $\hat{H}_\text{eff}^\sigma$. Alternatively, and this is done
in the present work, one may also determine the coefficients $a_a^{j\sigma}$ by solving (\ref{eq:tdev_dft}) as a generalized eigenvalue problem 
and making use of transformations
between the basis sets (\ref{Basis_diag}) and (\ref{Basis}) (see appendix~\ref{sec:trafos}).

In the basis~(\ref{Basis_diag}) and with the absorber~(\ref{eq:vabs2}) the derivative of the norm (\ref{eq:normaenderung_naqmd}) becomes
\begin{equation}
	\ddt N = -2 \sum_{\sigma=\uparrow,\downarrow} \sum_{j=1}^{N_\text{e}^\sigma} \sum_{a=1}^{N_\text{b}} V_{\text{abs, }aa}^\sigma \left| a_a^{j\sigma}\right|^2
\end{equation}
and the additional term~(\ref{Deltaabs}) of the energy balance~(\ref{EB_abs_me}) is
\begin{equation}
	\Delta_\text{abs}
	= 
	- 2\, \sum_{\sigma=\uparrow,\downarrow} \sum_{j=1}^{N_\text{e}^\sigma} \sum_{i=a}^{N_\text{b}}
		V_{\text{abs, }aa}^\sigma \epsilon_a^\sigma \left| a_a^{j\sigma}\right|^2  \,\text{.}
\end{equation}
Thus, both quantities are always zero or negative if
\begin{equation}
	V_{\text{abs, }aa}^\sigma = f_a^\sigma = \left\{
		\begin{array}{ccc}
			0 & \hspace{1cm}\text{for}\hspace{1cm} & \epsilon_a^\sigma \leq 0 \\
			\geq0 & \text{for} & \epsilon_a^\sigma > 0 
		\end{array}
	\right. \,.
\end{equation}
It now becomes apparent that our choice of the absorber potential does guarantee the physical requirement, namely, that density is removed
only from states which contribute to the continuum. The eigenvalues $f_a$ have still to be determined which will be done in the next section.

\subsection{The parameters of the absorber}
\label{sec:para}

The $f_a$ are directly connected to the lifetime $\tau_a$ of the states $|\chi_a\rangle$
if the Hamiltonian $\hat{H}$ is time-independent, i.e.
\begin{equation}
	f_a = +\frac{1}{2\tau_a} \,\text{.}
	\label{eq:def_f}
\end{equation}
We use here the quantities $\tau_a$ for an appropriate parameterization of $f_a$. It is natural to assume that the lifetimes increase smoothly with
energy. Thus, we parameterize the lifetimes according to
\begin{equation}
	\tau_a = 
	\left\{ 
		\begin{array}{ccc}
			\infty &\hspace{0.5cm}& \epsilon_a \leq 0\,\text{a.u.} \\
			\frac{\tau_\text{min}}{\sin^2\left(\frac{ \epsilon_a \pi}{2 E_\text{ref}} \right)}  &\hspace{0.5cm}& 0 < \epsilon_a < E_\text{ref} \\
			\tau_\text{min}  &\hspace{0.5cm}& \epsilon_a \geq E_\text{ref}
		\end{array}
	\right\}
	\label{eq:tau_i}
\end{equation}
where the $\epsilon_a$ are the ``field-following'' time-dependent energies~(\ref{eq:tdev_dft}) and
$\tau_\text{min}$ and $E_\text{ref}$ are two parameters still to be determined.

The general conditions that the parameters $\tau_\text{min}$ and $E_\text{ref}$ (or the absorbing potential at all) have to satisfy are similar to
those in calculations on spatial grids. In this case, the absorber has be strong enough to prevent unphysical reflections of the density at the boundaries of
the grid. On the other side, it must be weak enough to prevent reflections at the absorbing potential itself~\cite{Dundas_privatecomm}.
In our case, this is equivalent to the requirements, that the absorption is strong enough to prevent reflections at the boundary of the Hilbert space
(defined by the finite number of basis states). On the other side, it should be weak enough to avoid any suppression of excitation.

To illustrate this somewhat exceptional, quantum mechanical property one can consider a two level system 
\begin{eqnarray}
	\rmi\ddt a_{1}(t) &=& E_1(t) a_1(t) + H_{12}(t) a_2(t) \,,\\
	\rmi\ddt a_{2}(t) &=& E_2(t) a_2(t) + H_{12}(t) a_1(t)
\end{eqnarray}
where $a_{1/2}$ are the expansion coefficients of the two states, $E_{1/2}$ the energies and $H_{12}$ is the coupling matrix element.
It is assumed without loss of generality that the basis states $|\Phi_1\rangle$ and $|\Phi_2\rangle$ are orthogonal.
The population of the first state $\left| a_1 \right|^2$ is constant if the absorption in the second state is so strong that $a_2=0$ for all times, i.e.
\begin{equation}
	\ddt |a_1(t)|^2 = \rmi H_{12}(t)( a_2(t)^* a_1(t) - a_1(t)^* a_2(t) ) = 0 \hspace{0.5cm} \text{if} \hspace{0.5cm} a_2(t)=0 \hspace{0.2cm} \forall t \,.
\end{equation}
Obviously, in this case, the absorber completely prevents any excitation.

In the following, we use $E_\text{ref}=0.3$~a.u. because the underlying Hilbert space (i.e. the finite basis set used) yields a dense spectrum of states
up to this energy. This fulfilles the second requirement, discussed above. The minimal decay time, which determines the strength of the absorber, is fixed to
be $\tau_\text{min}=5$~a.u. In our test calculations we found, however, that a relatively large range of parameters leads to very similar results. In
addition, the above given and fixed parameters are applicable in a very large range of laser parameters as will be shown in the following by comparing the 
results with that of reference calculations.

\section{Results}
\label{sec:res}

\subsection{The atomic benchmark system: The hydrogen atom H}
\label{sec:H}

First, we test our approach on the hydrogen atom in intense laser fields. For this system it is possible to employ a very accurate description of
ionization without any absorbing boundary conditions using huge basis sets~\cite{Geltman_JPB_33_1967(2000),Hansen_PRA_64_033418(2001)}.
E.g., Hansen et al. have used 120 discrete and 2816 continuum states (build from hydrogen eigenfunctions with $m=0$) to investigate the laser induced dynamics
of H(1s)~\cite{Hansen_PRA_64_033418(2001)}.

\begin{figure*}[hbt]
	\begin{minipage}{0.48\textwidth}
		\mittig{without absorber}
		\mittig{\includegraphics[width=0.9\columnwidth]{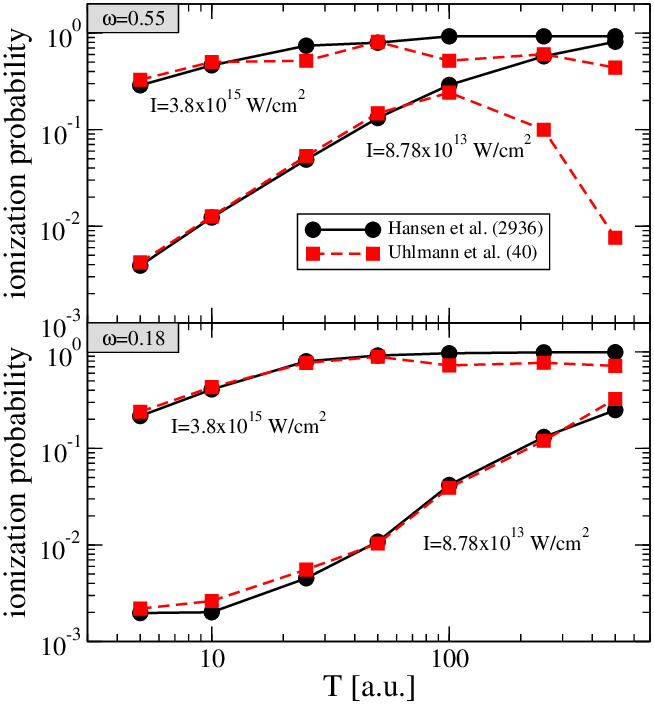}}
	\end{minipage}
	\begin{minipage}{0.02\textwidth}
		\hfill
	\end{minipage}
	\begin{minipage}{0.48\textwidth}
		\mittig{with absorber}
		\mittig{\includegraphics[width=0.9\columnwidth]{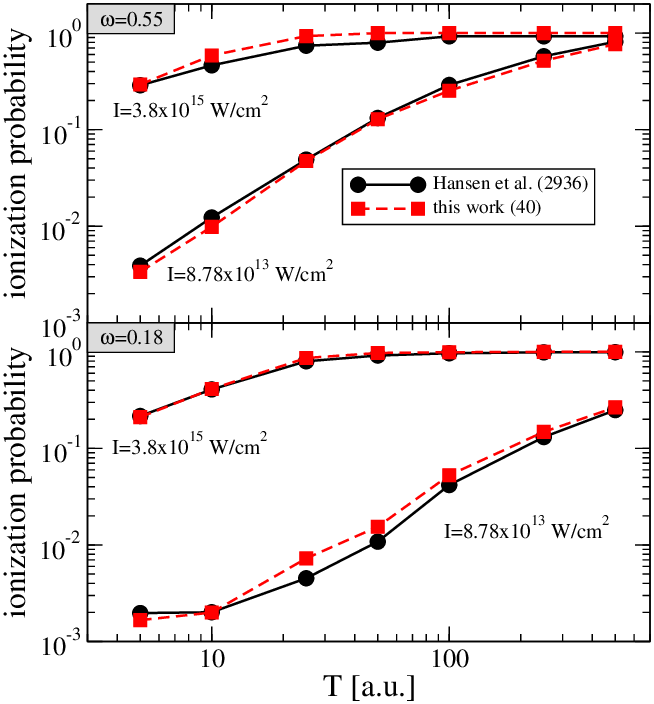}}
	\end{minipage}
	\caption{(Color online) The ionization probability of H(1s) as function of the laser pulse duration calculated without (left) and with (right) 
	absorber potential.
	Two frequencies ($\omega=0.55$~a.u. (top) and $\omega=0.18$~a.u. (bottom)) and two laser intensities 
	($8.78\times10^{13}\frac{\text{W}}{\text{cm}^2}$ (lower curves) and $3.8\times10^{15}\frac{\text{W}}{\text{cm}^2}$ (upper curves)) have been used. 
	Our results without absorber (Uhlmann et al.~\cite{BE1}) and with absorber (this work) are compared to the high precision data 
	of Hansen et al.~\cite{Hansen_PRA_64_033418(2001)}.
	 The numbers in brackets in the legends denote the size of the basis sets.}
	\label{fig:Vergleich_Hansen}
\end{figure*}

In contrast, we use here a basis of only 40 functions but include absorbing boundary conditions. 
The basis set contains the 1s, 2s and 2p$_z$ eigenfunctions of hydrogen extended
with chains of 37 Gaussians in the direction of the electric field of the laser~\cite{Uhlmann-Diplompaper}. The same basis has also been used in our previous
calculations of H(1s) in intense laser fields~\cite{BE1}, however, without absorber. 
It was found~\cite{BE1} that the initial excitation and ionization are described
very well but fail once the ionization probability becomes noticable (see left part of figure~\ref{fig:Vergleich_Hansen}).

In these and the present calculations, as well as in the reference calculations of Hansen et al.~\cite{Hansen_PRA_64_033418(2001)}
the hydrogen atom is exposed to a laser field of the form
\begin{equation}
	E(t) = E_0 f(t) \sin(\omega t + \varphi ) 
\end{equation}
with
the shape function $f(t)$
\begin{equation}
	f(t) = \left\{ 
		\begin{array}{ccc}
			\sin^2\left( \frac{\pi}{2T}t \right) &\hspace{2cm}& \text{for}\,\, 0 < t < 2T \\
			0 && \text{otherwise}
		\end{array}
		\right.
	\,\text{,}
	\label{eq:sin2}
\end{equation}
$T$ the duration of the laser pulse, $\omega$ the frequency, $\varphi$ the phase and $E_0$ the amplitude.
In our calculations without absorber~\cite{BE1} and in the calculations of Hansen et al.~\cite{Hansen_PRA_64_033418(2001)}
the ionization probability is defined as the part of the electronic
density that is in states with positive energy. In our present calculations with absorber the ionization probability is defined as 
\begin{equation}
	P_\text{ion} = 1 - N( t_\text{final} )
	\label{eq:Pion}
\end{equation}
where $N( t_\text{final} )$ is the norm of the wave function at the end of the calculation with $t_\text{final} = 2\,T + 500$~a.u. 
The additional time interval of 500~a.u. ensures that
the norm has definitely reached its plateau after the laser pulse.

In figure~\ref{fig:Vergleich_Hansen}, the ionization probabilities of H(1s) as a function of the laser pulse duration are shown. The results on the left/right
side of figure~\ref{fig:Vergleich_Hansen} have been obtained in calculations without/with absorber potential.
Two intensities and two wavelengths are considered. The high precision data of Hansen et al.~\cite{Hansen_PRA_64_033418(2001)} are included.
As mentioned already, good agreement between our calculations without absorber and those of Hansen et al.~\cite{Hansen_PRA_64_033418(2001)} is 
found only if the laser pulse is relatively short or weak, i.e. if the
ionization probability is small (left graphs of figure~\ref{fig:Vergleich_Hansen}). The agreement for
longer or more intense laser pulses is dramatically improved if the absorber is included (right graphs of figure~\ref{fig:Vergleich_Hansen}).

Thus, it is possible to use a small basis set together with absorbing boundary conditions
instead of an accurate treatment of the continuum.
This is of special importance for molecular many-electron calculations. 
For these systems it is essential to use small basis sets and, thus, to introduce absorbing boundary conditions.

\subsection{The molecular benchmark system: The aligned hydrogen molecular ion H$_2^+$}
\label{sec:H2}

Next, the absorber is tested on the molecular benchmark system, \emph{pre-aligned} H$_2^+$, i.e. the molecular axis is oriented
parallel to the electric field of the laser and, thus, the nuclear motion is restricted to this axis. 
For this system, full quantum mechanical calculations
have been performed by Chelkowski et al.~\cite{Chelkowski-PRA_52_2977(1995)} which can serve as reference calculations.

\begin{figure}[tbh]
	\begin{center}
		\includegraphics[width=0.8\columnwidth]{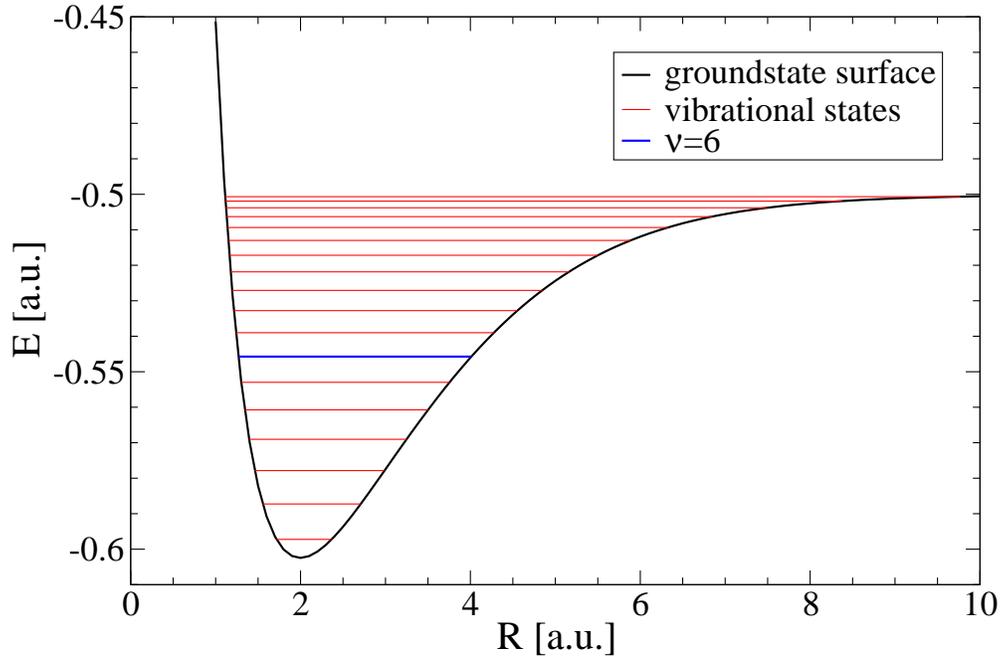}	
	\end{center}
	\caption{(Color online) Groundstate curve of H$_2^+$ (black) and vibrational levels (thin red lines, thick blue line for $\nu=6$)}
	\label{fig:gs_H2+_aligned}
\end{figure}

In contrast to our previous studies with a minimal~\cite{KS03} and extended~\cite{Uhlmann-Diplompaper} LCAO (linear combination of atomic orbitals)
basis we use here an elaborate basis set consisting of uncontracted Gaussians only (details are described in appendix~\ref{app:basish2+}).
It delivers an excellent description of the ground state surface with an equilibrium distance of $R_\text{eq}=1.9975$~a.u. and a minimum at
$E_\text{min}=-0.60246$~a.u. (see fig.~\ref{fig:gs_H2+_aligned}). The ground state energy $E_0 = E_\text{min}+\frac{\omega}{2}$ and the vibrational
levels $E_n$ ($n=1,2,\dots$) are also shown in fig.~\ref{fig:gs_H2+_aligned}. They are calculated according to the Bohr-Sommerfeld formula
\begin{equation}
	\oint p(R,E)\text{d}R = (n+n_0)\times h
	\label{eq:Bohr_Sommerfeld}
\end{equation}
with $n_0=1/2$ for the harmonic oscillator, $n$ the vibrational quantum number, $h$ the Plancks constant and $p(R,E)$ the classical
momentum as function of the energy $E$ and the distance $R$. This yields a binding energy of $E_0=-0.59724$~a.u., which is of 
quantum-chemical accuracy (cf. e.g.~\cite{Yan-PRA_67_062504(2003)}).

The H$_2^+$ is now exposed to the quasi-cw laser with the parameters according to that of Chelkowski et al.~\cite{Chelkowski-PRA_52_2977(1995)}. 
So, the laser has a short turn-on of 1~fs. 
The shape is kept constant afterwards. The frequency is $\omega=0.21$~a.u~$=5.71$~eV and the intensity is $3.5\times10^{13}$~W/cm$^2$. 
To obtain probabilities, 1000 trajectories were calculated and the results were averaged. The initial conditions of the trajectories were chosen 
according to the classical distance distribution in the 6th vibrationally excited state. Probabilities are defined as an average over the 
respective quantity for a single trajectory, i.e.
\begin{equation}
	P_{<\text{quantity}>} = \frac{1}{n} \sum_{i=1}^n P_{<\text{quantity}>}^i \,\text{.}
\end{equation}
The ionization probability for one trajectory $i$ is defined as the missing part of the norm, i.e.
\begin{equation}
	P_\text{ion}^i( t ) = 1 - N_i( t ) \,\text{.}
\end{equation}
The fragmentation probability is defined as
\begin{equation}
	P_\text{frag}^i( t ) = 
	\left\{ 
		\begin{array}{ccc}
			0 & \hspace{0.5cm} \text{for} & R( t ) < R_\text{D} \\
			1 & & \text{otherwise}
		\end{array}
	\right. 	
\end{equation}
with $R_\text{D}=9.5$~a.u. taken from~\cite{Chelkowski-PRA_52_2977(1995)}.
In accordance with Chelkowski et al. a dissociation probability, i.e. fragmentation without ionization, is defined as
\begin{equation}
	P_\text{diss}^i( t ) = ( 1 - P_\text{ion}^i( t ) ) \, P_\text{frag}^i( t ) \,\text{.}
\end{equation}

\begin{figure}[hbt]
	\begin{center}
		\includegraphics[width=0.8\columnwidth]{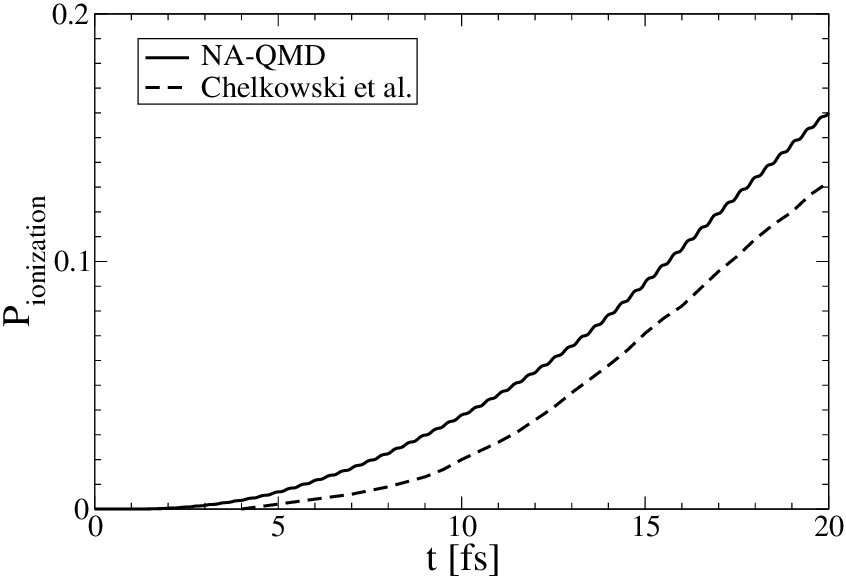}	
	\end{center}
	\begin{center}
		\includegraphics[width=0.8\columnwidth]{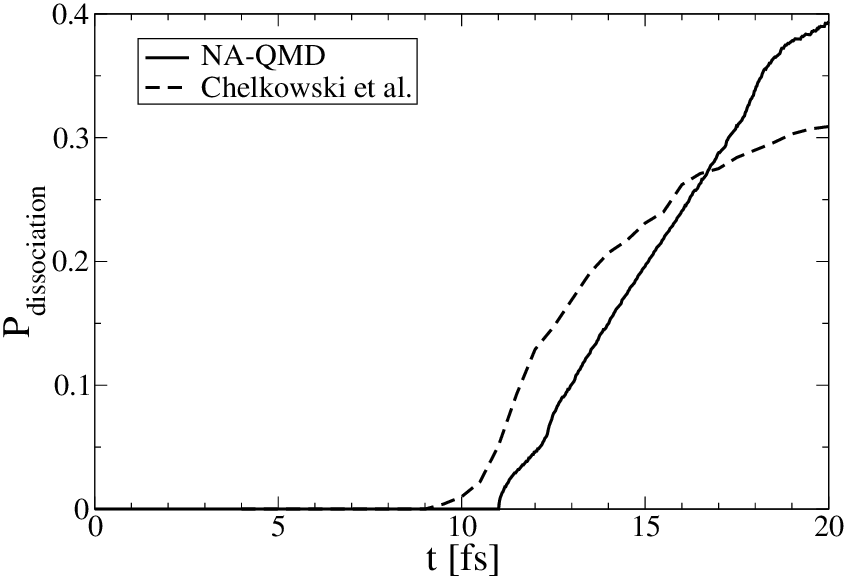}	
	\end{center}
	\caption{Ionization (top) and dissociation (bottom) probabilities 
	of H$_2^+$ as function of time obtained with the present NA-QMD method and by Chelkowski et al.~\cite{Chelkowski-PRA_52_2977(1995)}.
	(see text for details)}
	\label{fig:Pdion_H2+_aligned}
\end{figure}

The resulting ionization probability is shown in comparison to the full quantum mechanical results of
Chelkowski et al.~\cite{Chelkowski-PRA_52_2977(1995)} in the upper part of figure~\ref{fig:Pdion_H2+_aligned}.
The present calculations result in an ionization probability that is slightly higher than the full-quantum mechanical results. This is
due to the different definitions of the ionization probabilities.
Whereas in grid calculations electronic density can be counted as ionized only at large distances from the center of the system 
(e.g. outside a cylinder with $R<32$~a.u. in~\cite{Chelkowski-PRA_52_2977(1995)}), our absorber~(\ref{eq:vabs2}) acts also in the vicinity of the nuclei.
Therefore, the onset of ionization is somewhat
earlier in the present calculation as compared to the results of Chelkowski et al.~\cite{Chelkowski-PRA_52_2977(1995)}. After 12~fs both
ionization probabilities have a nearly linear slope. Regarding the uncertainties resulting from the different definitions of absorbing boundary
conditions, 
all in all, very good agreement between the present and the full quantum mechanical calculation is found.

In the lower part of figure~\ref{fig:Pdion_H2+_aligned}, the associated dissociation probabilities are shown. 
As seen, the onset of fragmentation is slightly delayed in the NA-QMD calculations as compared to the full quantum mechanical result. 
This is clearly due to the classical description of the nuclei.
Afterwards, the 
same behavior is observed, i.e. first a steep rise and after 18~fs the dissociation probability seems to run into a plateau. In spite
of this qualitative agreement, the present calculation overestimates the dissociation probability by 0.08.

\subsection{Non-aligned H$_2^+$ and H$_2$ molecules}
\label{sec:H2_2}

In this section, ionization probabilities and rates for H$_2^+$ and H$_2$ are considered as function of the angle $\Theta$ between the
molecular axis and the laser polarization axis as well as as function of the internuclear distance $R$ between the nuclei. Due to the lack
of unaligned studies, comparison with previous work can be done only for aligned molecules, i.e. 
$\Theta=0$~\cite{Yu-PRA_54_3290(1996),Saenz-PRA_61_051402(2000),Saenz-PRA_66_063407(2002),Harumiya-JCP_113_8953(2000),Harumiya-PRA_66_043403(2002),Lein-PRA_65_033403(2002)}.

All calculations have been performed with fixed nuclei. The H$_2$ molecule is considered in the TDHF approximation (see~\cite{Kunert-Ethylen} for details).
To account the orientation dependence the local basis functions centered at the nuclei are extended with functions located at hexagonal grid points
(see appendix~\ref{app:basish2} for details). With this basis the ground state surfaces exhibit equilibrium internuclear distances and energies of
$R_\text{eq}^{H_2^+}=1.99744$~a.u., $E_\text{min}^{H_2^+}=-0.602455$~a.u. and $R_\text{eq}^{H_2}=1.39384$~a.u., $E_\text{min}^{H_2}=-1.13608$~a.u.

\paragraph{Ionization probabilities}
In the following, the ionization probabilities of H$_2^+$ and H$_2$ as function of $\Theta$ will be discussed. They are calculated at the equlibrium internuclear
distance with a 50~fs, 266~nm and $5\times10^{14}$~W/cm$^2$ laser pulse.
For H$_2^+$, the ionization probability $P_\text{ion}$ is again defined as the missing part of the norm, eq.~(\ref{eq:Pion}). 
For H$_2$, we abbreviate the missing part of the norm of the single-particle functions with
\begin{equation}
	P_{\text{s,}1/2} = 1-N_{1/2}
\end{equation} 
where $N_{1/2}$ are the norms of the single-particle wave functions of the two electrons.
Because both electrons differ only by their spin degrees of freedom one has of course $N_1=N_2$ and $P_{\text{s,}1}=P_{\text{s,}2}=P_\text{s}$.
The single and 
double ionization probability are obtained via the single-particle approximation~\cite{Lappas-JPB_31_L249(1998),Petersilka-LP_9_105(1999)}
\begin{eqnarray}
	P_\text{single} &=& (1-N_1)N_2 + N_1(1-N_2) \label{eq:Pion_single} \,\text{,} \\
	P_\text{double} &=& (1-N_1)(1-N_2) \label{eq:Pion_double} \,\text{.}
\end{eqnarray}
Note, that with the definition~(\ref{eq:Pion_single}) the maximum of $P_\text{single}$ is $0.5$ if $N_1=N_2$.

\begin{figure}[hbt]
	\mittig{\includegraphics[width=0.8\columnwidth]{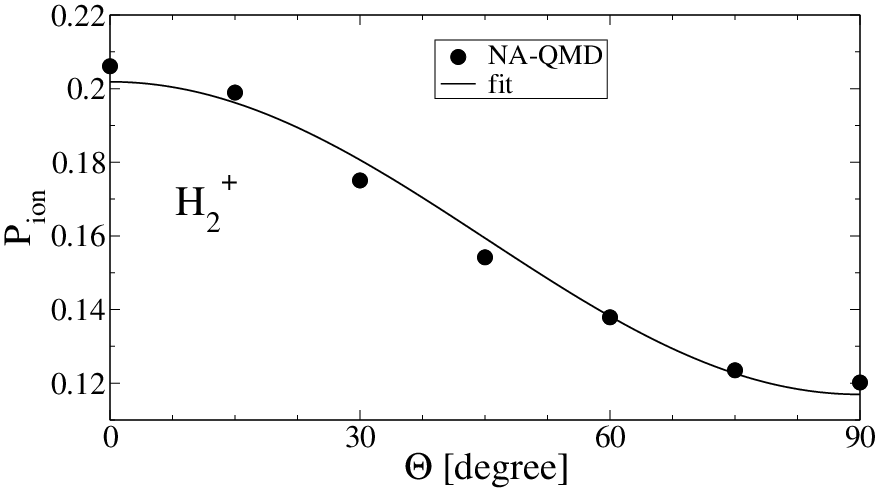}}
	\mittig{\includegraphics[width=0.8\columnwidth]{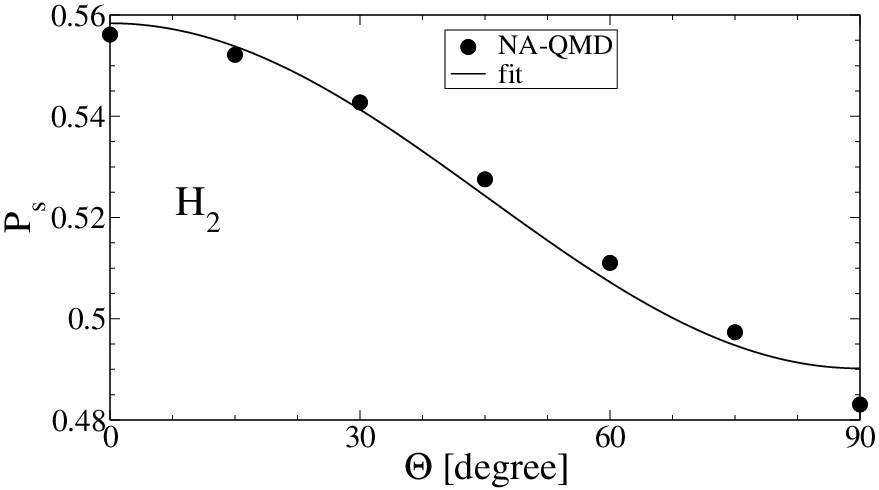}}
	\mittig{\includegraphics[width=0.8\columnwidth]{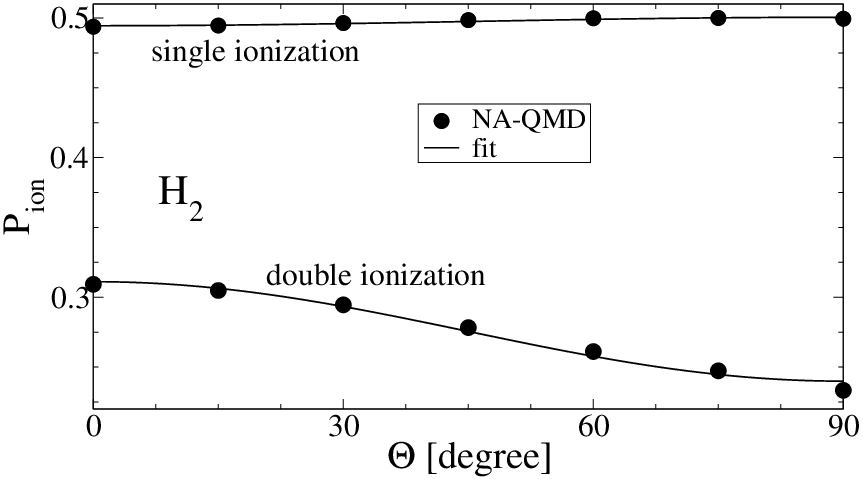}}
	\caption{
	Ionization probabilities $P_\text{ion}$ for H$_2^+$ and H$_2$ (top and bottom). Also, the missing part
	of the norm of the single-particle wave functions $P_{\text{s,}1}=P_{\text{s,}2}=P_\text{s}$ in H$_2$
	is shown (middle).
	The symbols denote the results of the
	NA-QMD calculation, the full lines the $P_\parallel \,\cos^2\Theta + P_\perp \, \sin^2\Theta$ fit (see text).}
	\label{fig:finorms_H2_H2+}
\end{figure}

$P_\text{ion}$, P$_\text{s}$, $P_\text{single}$ and $P_\text{double}$ are plotted as function of the angle $\Theta$ in fig.~\ref{fig:finorms_H2_H2+}.
All calculated probabilities have been fitted according to
\begin{equation}
	\Pi( \Theta ) = \left( P_\parallel \,\cos^2\Theta + P_\perp \,\sin^2\Theta \right) \,\text{.}
	\label{eq:fit}
\end{equation}
This parameterization has been used by Litvinyuk et al.~\cite{Litvinyuk-PRL_90_233003(2003)} to fit their experimental data on N$_2$~\cite{Litvinyuk-PRL_90_233003(2003)}.
The parameterization~(\ref{eq:fit}) fails to fit the experimental data on O$_2$~\cite{Alnaser-PRL_93_113003(2004)}. As can be seen from fig.~\ref{fig:finorms_H2_H2+},
the parameterization~(\ref{eq:fit}) works very well for all probabilities in H$_2^+$ and H$_2$ where experimental data do not exist to date.

From fig.~\ref{fig:finorms_H2_H2+} it becomes apparent that the probabilities $P_\text{ion}$ and $P_\text{s}$ are largest at parallel orientation $\Theta=0$~degree
and decrease with increasing $\Theta$ exhibiting the minimum at perpendicular orientation $\Theta=90$~degree, for both molecules, as expected.
However, the orientations dependence of the response is much more pronounced for H$_2^+$ as compared to H$_2$ (note the different absolute values of the 
probabilities in the upper and middle part of fig.~\ref{fig:finorms_H2_H2+}). In addition, the single ionization probability in H$_2$ is practically
orientation independent, at least for the chosen laser parameters where the double ionization probability is not small (lower part of fig.~\ref{fig:finorms_H2_H2+}).

Striking differences between H$_2^+$ and H$_2$ have been found also in the alignment behavior of both molecules~\cite{Uhlmann-Alignment}. In a relatively large
range of laser intensities, fragments originating from H$_2^+$ are much more aligned as compared to those from H$_2$~\cite{Uhlmann-Alignment}. This is
in accord with the present findings of the much larger anisotropic response of H$_2^+$ in comparison to H$_2$.

\paragraph{Ionization rates}

\begin{figure}[hbt]
	\mittig{\includegraphics[width=0.9\columnwidth]{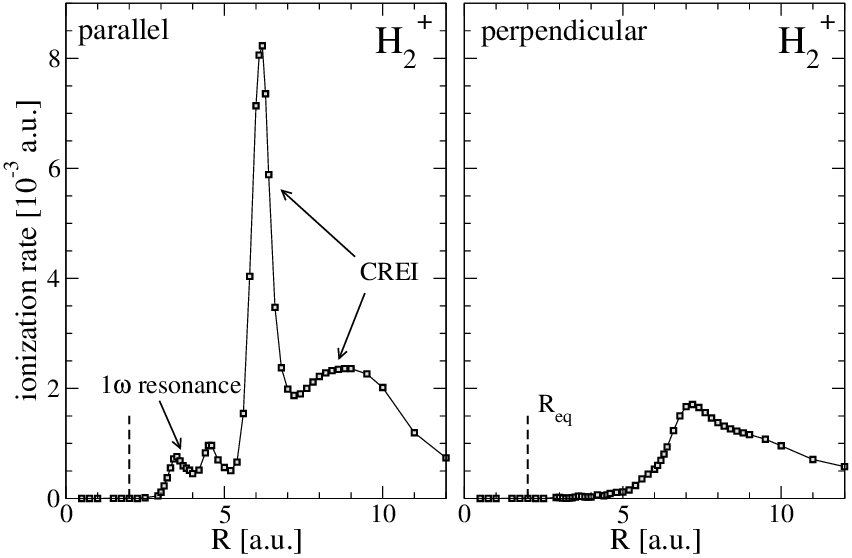}}
	\mittig{\includegraphics[width=0.9\columnwidth]{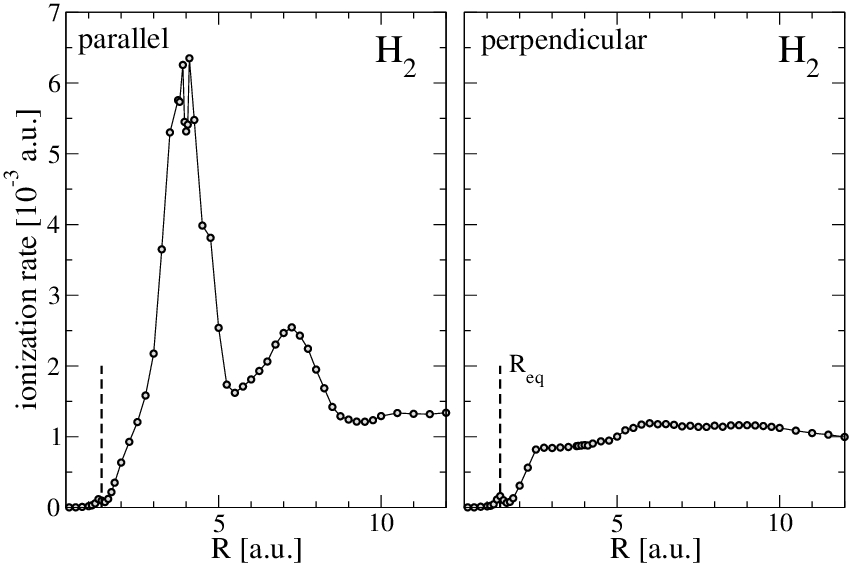}}
	\caption{Ionization rates as function of the internuclear distance for H$_2^+$ (top) and H$_2$ (bottom) at a laser wavelength of 266~nm 
	and an intensity of $8\times10^{13}$~W/cm$^2$. The vertical broken line indicates the equilibrium internuclear distance.}
	\label{fig:ionraten_model_H2}
\end{figure}

Finally, the ionization rates $\Gamma( s^{-1})$ as function of the internuclear distance for parallel and perpendicular orientations will be considered for both molecules.
They are calculated from the logarithmic decrease of the norm $N(t)$, i.e. 
\begin{equation}
	\text{ln} N(t) = - \Gamma \,t \,.
\end{equation}
In the case of H$_2$ one has $\Gamma=\Gamma_1 + \Gamma_2$ in the spirit of the independent particle model (i.e. $N(t)=N_1(t)\,N_2(t)$).
The cw-laser used has a short turn-on of three optical cycles and a constant shape afterwards. The wavelength is 266~nm and different intensities have been used.
For the intensity of $8\times10^{13}$~W/cm$^2$ the resulting ionization rates as function of the internuclear distance $R$ are shown in fig.~\ref{fig:ionraten_model_H2}
for parallel and perpendicular orientation and both molecules.

For H$_2^+$ and parallel orientation (left upper part of fig.~\ref{fig:ionraten_model_H2}), the well-known features are recovered with our new formalism, i.e.,
enhanced ionization is observed for internuclear distances between 6 and 10~a.u.
This is the well-known charge-resonance enhanced
ionization (CREI)~\cite{Zuo-PRA_52_R2511(1995),Seideman-PRL_75_2819(1995)} which is accompanied by an electron 
localization~\cite{Seideman-PRL_75_2819(1995),Posthumus-JPB_28_L349(1995),Posthumus-JPB_29_5811(1996)}.
The CREI features are observed here although the system is in the multi photon regime, i.e. the Keldysh parameter (see e.g.~\cite{Dundas-PRA_71_013421(2005)})
is $\gamma\gg1$. This emphasizes the generality of the CREI mechanism.
At $R\approx3.5$~a.u the one-photon resonance between the electronic groundstate and the first excited state
manifests itself as a peak in the ionization rates in parallel orientation. 
An additional peak is found at $R\approx4.5$~a.u. This peak is suppressed for higher intensities, for which the results are not included 
in figure~\ref{fig:ionraten_model_H2}.

For H$_2$ and parallel orientation (left lower part of fig.~\ref{fig:ionraten_model_H2}), the ionization rate exhibits
two peaks which look similar to the CREI peaks
of H$_2^+$ in parallel orientation. These peaks are located at smaller internuclear distances than for H$_2^+$.
In addition, one of the peaks is split which might originate from resonances as the system is in the multiphoton regime.

The occurrence of enhanced ionization in H$_2$ has been also subject to a number of previous studies. 
First, the distance dependent ionization of H$_2$ was investigated using a 1D model~\cite{Yu-PRA_54_3290(1996)}. Depending
on laser intensity and frequency either one or two peaks were found. E.g., at $\lambda=532$~nm and $I=1\times10^{14}$~W/cm$^2$ enhanced ionization (EI)
was found at $R\approx4$~a.u. and $R\approx6$~a.u. which is very similar to the findings in our fully 3D calculations.
The behavior of parallel aligned H$_2$ in static electric fields was studied with high precision quantum chemistry
methods~\cite{Saenz-PRA_61_051402(2000),Saenz-PRA_66_063407(2002)}, too. An avoided crossing of the H$_2$ groundstate and an excited state,
that corresponds to the ionic fragments H$^+$ and H$^-$ in the static electric field, was found.
In a fully 3D study of parallel aligned H$_2$~\cite{Harumiya-JCP_113_8953(2000),Harumiya-PRA_66_043403(2002)}
it was found as well that the enhanced ionization is linked to the formation of ionic components H$^+$ and H$^-$ as a typical signature of CREI in the H$_2$ molecule.

We note also, that the direct experimental observation of CREI in H$_2^+$ has been published recently~\cite{Pavicic-PRL_94_163002(2005)}.
For H$_2$ this is not the case. Moreover, it has been shown recently in a 1D study~\cite{Lein-PRA_65_033403(2002)} that ionization of H$_2$
usually takes place near the equilibrium internuclear distance. Thus, the role of the CREI mechanism in aligned H$_2$ 
remains still an open problem for future investigations.

The most surprising result of the present studies concerns the ionization probabilities in perpendicular orientation
(right part of fig.~\ref{fig:ionraten_model_H2}). Clearly seen, the calculations predict enhanced ionization for both
molecules. In H$_2^+$, the ionization rate exhibits a distinct maximum at $R\approx7$~a.u., whereas in H$_2$, this quantity clearly
increases around $R\approx2.5$~a.u. (Note also, that tiny peaks are found in H$_2$ near the equilibrium distance $R_\text{eq}$ for both
orientations, and cf. discussion above). One can definitely exclude the CREI mechanism to be responsible for the enhanced ionization in
perpendicular orientation, because electron localization cannot take place in this geometry.
In H$_2^+$, this feature probably originates from a resonance that occurs around $R\approx7$~a.u. However, this as well as the nearly
structureless enhancement of the ionization rate in H$_2$ remain as subjects for systematic future studies.

\section{Conclusions}
\label{sec:concl}

It was the main aim of this work to present, for the first time, a method to introduce absorbing boundary conditions in calculations of many electron systems
with basis expansion. The basic idea is rather general (section~\ref{sec:abs_oe}) and based on an imaginary potential
constructed as a projection operator with time-dependent adiabatic states. In this paper it has been used to extend the NA-QMD formalism in order to
describe realistically ionization processes in many electron systems.

The method was tested on the benchmark systems, the hydrogen atom and the aligned H$_2^+$ molecule in intense laser fields. Very good agreement 
between the calculated ionization probabilities and that of (numerically very extensive) reference calculations has been found.

The extended NA-QMD formalism allowed us, also for the first time, to study the ionization process in unaligned H$_2^+$ and H$_2$ molecules. A
completely different orientation dependence of the ionization probabilities in both molecules was found, with a distinct anisotropic
response in H$_2^+$ and a nearly isotropic behavior in H$_2$. In addition, enhanced ionization for perpendicular orientation in
both molecules is predicted, which to the best of our knowledge, has not been reported before.

The present method can be applied to larger systems, like N$_2$~\cite{Litvinyuk-PRL_90_233003(2003)}, 
O$_2$~\cite{Alnaser-PRL_93_113003(2004)} or organic molecules where
the experimental findings still require a consistent interpretation.
Finally, we note also that the present method can also be used to study ionization processes in atomic collisions with many electrons~\cite{Luedde-Buch}.

\begin{acknowledgments}
We thank 
the Deutsche Forschungsgemeinschaft (DFG) for support through grants SCHM~957/6-1 and GR~1210/3-3 and the 
Zentrum f\"ur Hochleistungsrechnen (ZHR) of the Technische Universit\"at Dresden for providing computation time.
\end{acknowledgments}

\appendix
\section{Basis transformations}
\label{sec:trafos}
In order to solve the time-dependent eigenvalue problem~(\ref{eq:tdev_dft})
\begin{equation}
	\hat{H}_\text{eff}^\sigma(t) |\chi_a^\sigma\rangle(t) = \epsilon_a^\sigma(t)  |\chi_a^\sigma\rangle(t)
	\label{eq:tdev_dft_app}
\end{equation}
the effective single-particle ``field-following'' adiabatic states are expanded in the basis $\{|\phi_\al\rangle\}$~(\ref{Basis})
\begin{equation}
	|\chi_a^\sigma\rangle = \sum_{\be=1}^{N_\text{b}} U_{a\be}^\sigma |\phi_\be\rangle \,.
	\label{eq:trafo_to_eb}
\end{equation}
Multiplying $|\phi_\al\rangle$ with $\sum_a|\chi_a^\sigma\rangle\langle\chi_a^\sigma|$ the expansion of $|\phi_\al\rangle$ in the basis $\{|\chi_a^\sigma\rangle\}$ results 
\begin{equation}
	|\phi_\al\rangle = \sum_{a\beta}^{N_\text{b}} S_\albe U_{\beta a}^{\sigma+} |\chi_a^\sigma\rangle \,.
	\label{eq:trafo_to_eb_back}
\end{equation}
Note, that $\sum_a|\chi_a^\sigma\rangle\langle\chi_a^\sigma|$ can be used like an identity because the basis sets $\{|\phi_\al\rangle\}$ and 
$\{|\chi_a\rangle\}$ span exactly the same part of the Hilbert space. Furthermore, the property
\begin{equation}
	\sum_{a=1}^{N_\text{b}} U_{\beta a}^{\sigma+} U_{a\gamma}^\sigma = \left( S^{-1} \right)_{\bega}
\end{equation}
is obtained by using (\ref{eq:trafo_to_eb}) and (\ref{eq:trafo_to_eb_back}).
Therefore, the transformation $\hat{U}^\sigma$ is unitary 
only if both basis sets are orthogonal, i.e. if also $\left( S^{-1} \right)_{\bega}=\delta_\bega$.

Inserting~(\ref{eq:trafo_to_eb}) into~(\ref{eq:tdev_dft_app}) and multiplying with $\langle\phi_\al|$ results in the generalized eigenvalue problem
\begin{equation}
	\sum_{\be=1}^{N_\text{b}} \left( 
		\langle \phi_\al | \hat{H}_\text{eff}^\sigma | \phi_\be \rangle - \epsilon_a^\sigma \langle \phi_\al | \phi_\be \rangle
	\right)  U_{a\be}^\sigma = 0 \,.
	\label{eq:gev}
\end{equation}
The effective single-particle energies $\epsilon_a^\sigma$ and the transformation $\hat{U}^\sigma$ are obtained by solving~(\ref{eq:gev}).

The effective single-particle wave function $|\Psi^{j\sigma}\rangle$~(\ref{Basis}) can either be expanded in the basis $\{|\phi_\al\rangle\}$ or
in the basis $\{|\chi_a\rangle\}$ 
\begin{equation}
	\label{Basis_app} |\Psi^{j\sigma}\rangle(t)
	=\sum_{\alpha=1}^{N_\text{b}} a^{j\sigma}_{\alpha}(t)|\phi_\alpha\rangle(t)) 
	=\sum_{a=1}^{N_\text{b}} a^{j\sigma}_{a}(t)|\chi_a\rangle(t)) 
	\hspace{1cm} \text{with} \hspace{0.5cm} j=1\dots N_\text{e}^\sigma,\,\sigma=\uparrow,\downarrow \,.
\end{equation}
Using equations~(\ref{eq:trafo_to_eb}) and (\ref{Basis_app})
the transformations for the coefficients $a_\alpha^{j\sigma}$ and $a_a^{j\sigma}$ 
\begin{eqnarray}
	a_a^{j\sigma} &=& \langle \chi_a^\sigma | \Psi^{j\sigma} \rangle = \sum_\albe^{N_\text{b}} U_{\al a}^{\sigma+} a_\be^{j\sigma} S_\albe \,,\\
	a_\al ^{j\sigma} &=& \sum_{a=1}^{N_\text{b}} U_{a\al}^\sigma a_a^{j\sigma}
\end{eqnarray}
and a general matrix $O_\albe^\sigma$ and $O_{ab}^\sigma$
\begin{eqnarray}
	O_{ab}^\sigma &=& \langle\chi_a| \hat{O} | \chi_b \rangle = \sum_\albe^{N_\text{b}} U_{\al a}^{\sigma+} O_\albe^\sigma U_{b\be}^\sigma \,,\\
	O_\albe^\sigma &=& \langle\phi_\al| \hat{O} | \phi_\be \rangle = 
		\sum_{ab\gamma\delta}^{N_\text{b}} S_{\al\gamma} U_{a\gamma}^\sigma O_{ab}^\sigma U_{\delta b}^{\sigma+} S_{\delta\beta}
\end{eqnarray}
are obtained. The last transformation is used to calculate the matrix elements $V_{\text{abs, }\albe}^\sigma$ which are used in practical calculations.

\section{Basis sets}

In this appendix all details of the basis sets used in the H$_2^+$ and H$_2$ calculations are given to make the calculations comprehensible.
The basis set used in the H calculations (section~\ref{sec:H}) consists of the hydrogen 1s, 2s and 2p$_\text{z}$ basis functions
extended with chains of s-type Gaussians and is described in detail in \cite{BE1}.

\subsection{Aligned H$_2^+$}
\label{app:basish2+}

The basis set that is used in the calculations to aligned H$_2^+$ (section~\ref{sec:H2}) is a combination of a local basis centered at each of the two nuclei and a chain of additional 
s-type Gaussians located along the laser polarization axis. Gaussians
\begin{equation}
	\phi_{Ailm}( \vec{r}\,' ) = N \, Y_{lm}( \theta', \phi' ) \, \text{e}^{-\frac{r_\text{A}^2}{\sigma_i^2}}
	\label{eq:allgGausse}
\end{equation}
are used also as basis functions at the nuclei.
In (\ref{eq:allgGausse}) $Y_{lm}$ are the spherical harmonics, $N$ is a norm constant, $\sigma_i$, $l$ and $m$ are parameters of the basis functions and
$\vec{r}_\text{A} = \vec{r} - \vec{R}_A$ where $\vec{R}_A$ is the center of the basis function (see e.g.~\cite{Kunert_Diss}).
At each of the two nuclei a basis set that consists of such Gaussians is located. 
The $\sigma_i$ are determined with
\begin{equation}
	\sigma_i = \sigma_1 \, f^{i-1} \hspace{2cm} (1\leq i \leq N)\,
	\label{eq:et_basis}
\end{equation}
where $\sigma_1$, $f$, $N$ are parameters given in table~\ref{tab:basis_H}.
With this basis set the atomic as well as the molecular groundstate are described extremely well. 
Furthermore, a good description of the excited states of H$_2^+$ is achieved with this basis set.
Please note, that the $\sigma_1$ for $l=1$ and $l=2$ have been chosen in such a way that 
$\sigma_\text{max}^{l=2}=1.7^{1/3}\,\sigma_\text{max}^{l=1}=1.7^{2/3}\,\sigma_\text{max}^{l=0}$.

\begin{table}[hbt]
	\begin{ruledtabular}
		\begin{tabular}{c|c|c|c|c}
			l & f & $\sigma_1$ [a.u.] & $\sigma_\text{max}$ [a.u.] & N \\
			\hline
			0 & 1.7 & 0.05 & 3.487 & 9 \\
			1 & 1.7 & 0.8473 & 4.162 & 4 \\
			2 & 1.7 & 1.7191 & 4.968 & 3 
		\end{tabular}
	\end{ruledtabular}
	\caption{Gaussian basis centered at each of the protons of H$_2^+$. The parameters given are those of equation~(\ref{eq:et_basis}).}
	\label{tab:basis_H}
\end{table}

The parameters of the additional chain of s-type Gaussians (see e.g.~\cite{BE1,Uhlmann-Diplompaper}) that is laid out symmetrically to the 
origin along the $z$-axis are given in
table~\ref{tab:basis1c_chain_old}. These additional functions have nearly no influence on the already excellent description of the groundstate and
the lowest excited states. They do, however, improve the description of highly excited and ionized electronic states 
and a dense level structure around $E=0$ results. The parameters given in table~~\ref{tab:basis1c_chain_old} were determined using the formalism
described in~\cite{BE1}.

\begin{table}[hbt]
	\begin{ruledtabular}
		\begin{tabular}{c|c|c}
		$\sigma$ [a.u.] & $d$ [a.u.] & $n$ \\
		\hline
		5.54 & 3.7 & 21
		\end{tabular}
	\end{ruledtabular}
	\caption{Parameters of the chain (width $\sigma$, spacing $d$ between neighboring functions and number $n$ of functions) 
	of s-type Gaussians laid out symmetrically to the origin along the $z$ axis.}
	\label{tab:basis1c_chain_old}
\end{table}

\subsection{Unaligned H$_2^+$ and H$_2$}
\label{app:basish2}

The same basis as before (see table~\ref{tab:basis_H}) is located at each of the hydrogen atoms in the calculations to unaligned H$_2^+$ and H$_2$ 
(section~\ref{sec:H2_2}).
However, a hexagonal grid of additional basis functions in the $y$-$z$ plane 
\begin{equation}
	\left( 
	\begin{array}{c}
	x_{ij}\\
	y_{ij}\\
	z_{ij}
	\end{array}
	\right) 
	=
	\left( 
	\begin{array}{c}
	0 \\
	\left( i - \frac{N_1}{2} \right) \frac{d}{2} \\
	\left( j - \frac{N_2}{2} \right) \sqrt{3} d + 
	\left| \left( i - \frac{N_1}{2} \right) \text{mod} 2 \right| \frac{\sqrt3}{2}d
	\end{array}
	\right) 
	\label{eq:hex}
\end{equation}
with
\begin{eqnarray}
	0 \leq i < N_1 && \text{and} \nonumber\\
	0 \leq j < N_2 - 1 & \hspace{0.2cm}\text{if } N_1+i \text{ even} & \text{or} \nonumber\\
	0 \leq j < N_2 & \hspace{0.2cm}\text{if } N_1+i \text{ odd} \nonumber
\end{eqnarray}
is used to extend the basis sets at the nuclei instead of the chain of Gaussians described in appendix~\ref{app:basish2+}. 
It is thus possible to
calculate the response of unaligned molecules to the laser field.
The parameters of the two hexagonal grids and the one chain are given in table~\ref{tab:basis3c_hex} and were also determined using
the formalism described in~\cite{BE1}.
The last set of Gaussians positioned at different places in space has such a large width and therefore also a large spacing that
it is not necessary to build a hexagonal grid for these Gaussians. Instead these basis functions are again laid out chain-like along
the $z$ axis~\cite{Uhlmann-Diplompaper}. 

\begin{table}[hbt]
	\begin{ruledtabular}
		\begin{tabular}{c|c|c|c}
		$\sigma$ [a.u.]  & d [a.u.] & $N_1$ & $N_2$ \\
		\hline
		5.74 & 5.2 & 9 & 7 \\
		7.81 & 10.38 & 5 & 3 \\
		16.62 & 18.68 & - & 3 
		\end{tabular}
	\end{ruledtabular}
	\caption{Parameters of the hexagonal grids (first two lines) and chain (bottom line) of s-type Gaussians laid out in the $y$-$z$ plane.}
	\label{tab:basis3c_hex}
\end{table}

It is reasonable to use a basis set for the density in the H$_2$ calculations. The ``exact'' density, i.e. the sum over the absolute square of the 
single-particle functions, is
expanded in this density basis to accelerate the calculation of Coulomb and xc matrix elements. 
Please note, that the norm of the density basis functions is different from the usual L$_2$({\bf R}$^3$) norm (see~\cite{Kunert_Diss}).
The parameters of the density basis used in the H$_2$ calculations are given in table~\ref{tab:basis_H_dens}.
$\sigma_1=0.05/\sqrt{2}$ has been chosen because the multiplication of two Gaussians with the width $0.05$ at the same place results in a Gaussian
with a width $0.05/\sqrt{2}$.
This density basis had not been transformed, i.e. the uncontracted Gaussians were used.
\begin{table}[hbt]
	\begin{ruledtabular}
		\begin{tabular}{c|c|c|c|c}
			l & f & $\sigma_1$ [a.u.] & $\sigma_\text{max}$ [a.u.] & N \\
			\hline
			0 & 1.7 & $\frac{0.05}{\sqrt{2}}=0.035355339$ & 2.46 & 9
		\end{tabular}
	\end{ruledtabular}
	\caption{Gaussian basis used as the H density basis. The parameters given are those of equation~(\ref{eq:et_basis}).}
	\label{tab:basis_H_dens}
\end{table}
Density basis functions are also located at the additional centers specified in table~\ref{tab:basis3c_hex}. The $\sigma_\text{dens}$ of the density
basis functions are set to $\sigma/\sqrt{2}$.

\bibliography{abs}

\end{document}